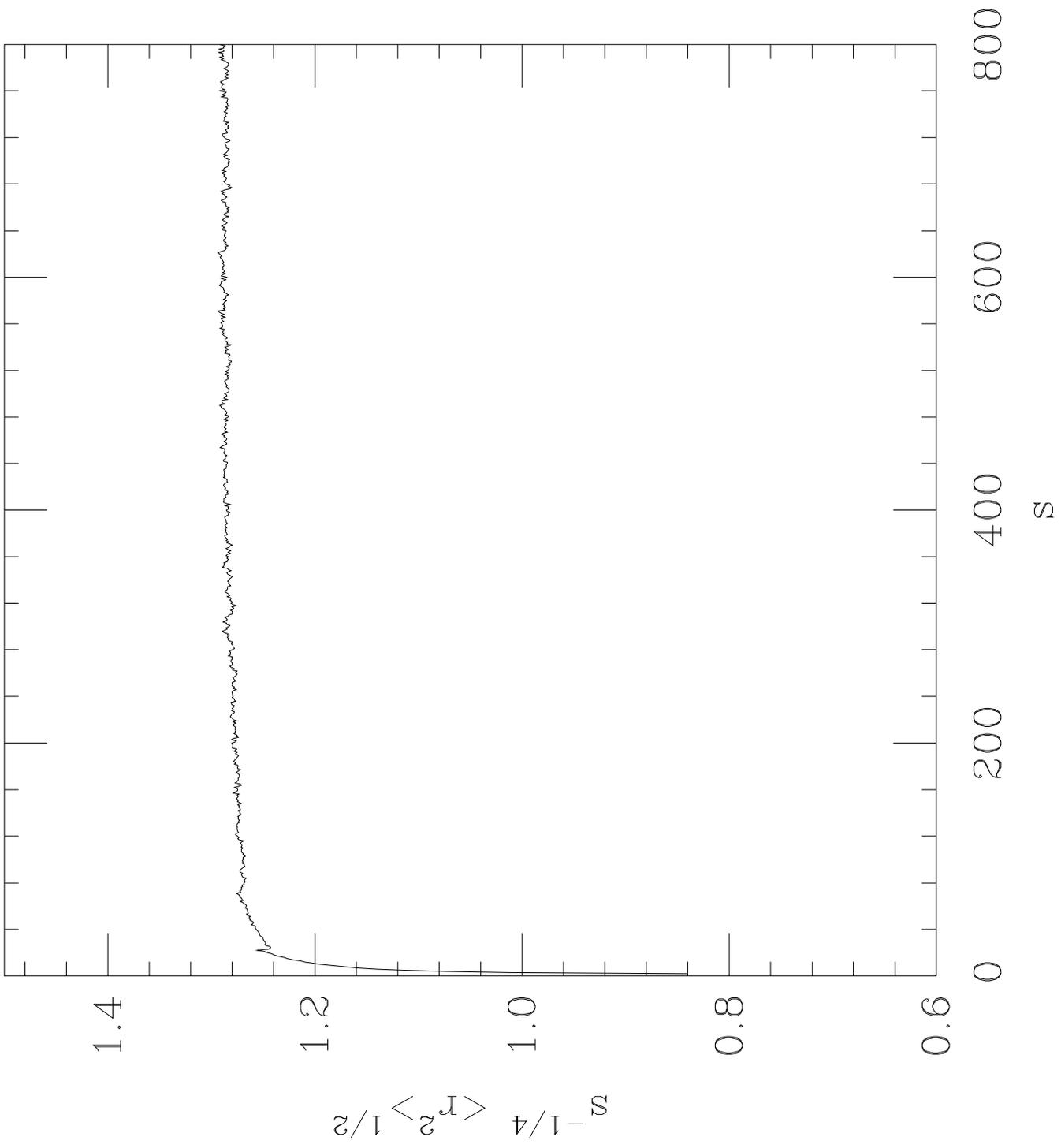

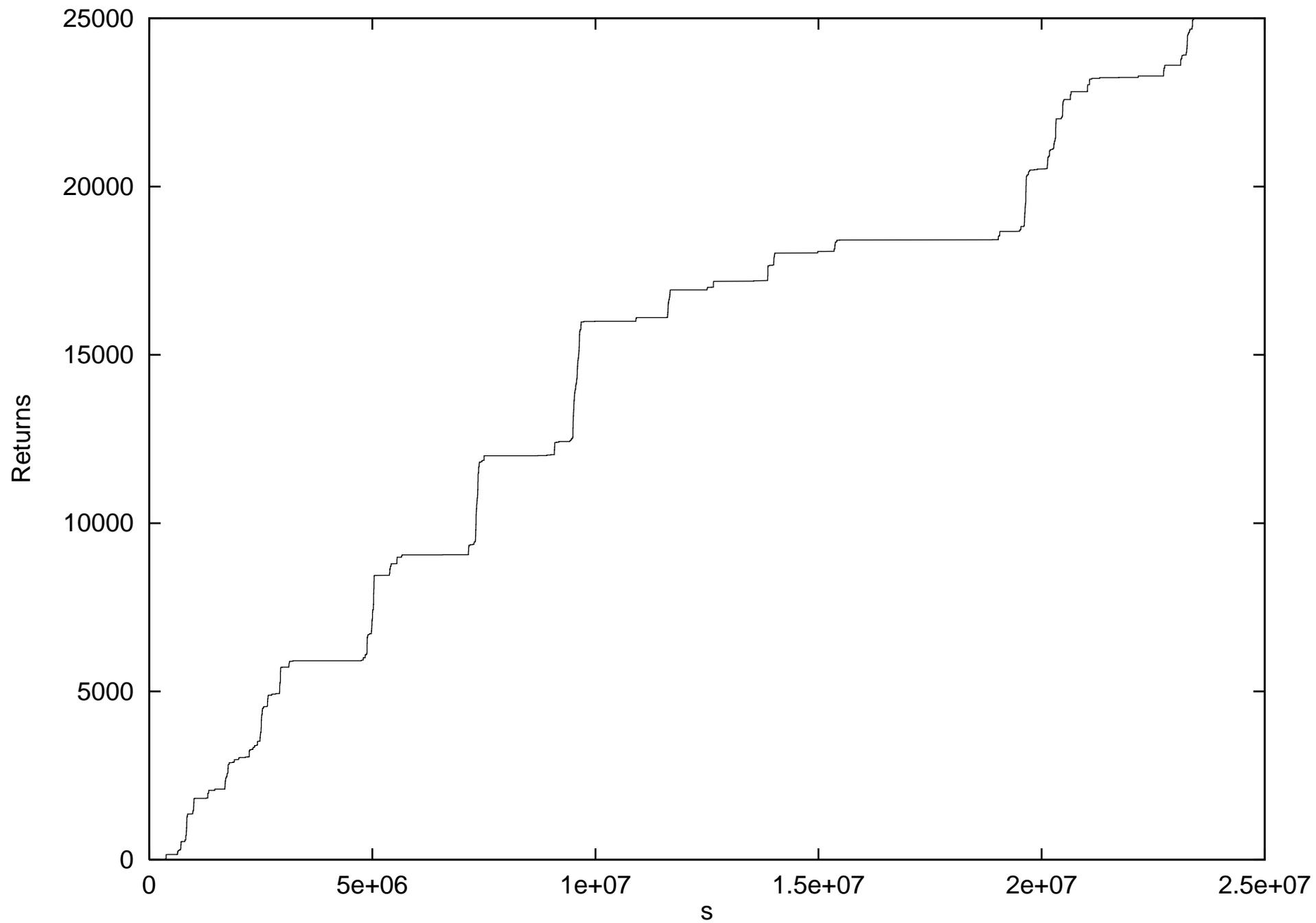

# Exact Results for Spatio-Temporal Correlations in a Self-Organized Critical Model of Punctuated Equilibrium

Stefan Boettcher and Maya Paczuski
*Department of Physics,*
*Brookhaven National Laboratory,*
*Upton, NY 11973*

November 20, 1995

We introduce a self-organized critical model of punctuated equilibrium with many internal degrees of freedom ($M$) per site. We find exact solutions for $M \to \infty$ of cascade equations describing avalanche dynamics in the steady state. This proves the existence of simple power laws with critical exponents that verify general scaling relations for nonequilibrium phenomena. Punctuated equilibrium is described by a Devil's staircase with a characteristic exponent, $\tau_{FIRST} = 2 - d/4$ where $d$ is the spatial dimension.

Many systems in nature, such as earthquake zones [1] or sand piles [2,3] are driven by an external force out of equilibrium into a highly correlated, critical state. Consecutive metastable states that the system inhabits are punctuated by avalanches, which dissipate the accumulated stress. These intermittent bursts eventually may be correlated over all sizes, indicating scale-free dynamics. This picture of self-organized criticality (SOC) [2] has led to intense studies of the dynamics of nonequilibrium systems. Much insight has been gained from the numerical investigation of models, for example, for invasion percolation [4], flux creep [5], depinning in quenched random media [6], biological evolution [7], and earthquakes [8]. A scaling theory has been developed for this broad range of models which is based on a few exact results [9] together with a scaling ansatz [10]. In addition, a mean-field theory has been proposed for the infinite range, random-neighbor evolution model [11,12]. For systems with punctuated equilibrium, though, the existence of simple power laws in the critical state has not been proven. Also, one would like to verify the scaling relations based on microscopic considerations, as is possible in equilibrium systems.

Here, we introduce a SOC model of punctuated equilibrium which is similar to that proposed by Bak and Sneppen in the context of biological evolution [7]. Our model specifies simple rules that may be plausible for a coarse grained description of evolution at the longest time scales, yet yields analytical results for robust features such as punctuated equilibrium which has been observed in the fossil record [13], and recently also in earthquake data [8]. Our main results are: (1) From microscopic dynamical considerations we derive equations of motion for the macroscopic observables. (2) We solve these non-linear equations of motion exactly, finding power laws with specific scaling coefficients. Punctuated equilibrium is described by a Devil's staircase with dimension-dependent, non-mean-field behavior. (3) These results verify general scaling relations for avalanche dynamics in systems out of equilibrium.

In our model, evolutionary activity is simulated in terms of mutation of the "least fit" species and interdependencies in a food chain, much like in the original Bak-Sneppen model. In addition, we consider the survivability of each species to be conditioned upon a number ($M$) of independent traits associated with the different tasks that it has to perform [14]. Our model is defined as follows: A species is represented by a single site on a lattice. The collection of traits for each species is represented by a set of $M$ numbers in the unit interval. A larger number represents a better ability to perform that particular task, while smaller numbers pose less of a barrier against mutation. Therefore, we "mutate" at every time step the smallest number among all species and among all traits. This number is replaced by a new number that is randomly drawn from a flat distribution in the unit interval, $\mathcal{P}$. The impact of this event on neighboring species is simulated by also replacing one of the $M$ numbers on each neighboring site with a new random number drawn from $\mathcal{P}$. Which one of the $M$ numbers is selected for such an update is determined at random since a mutation in the traits of one species can lead to adaptive change in any one of the traits of an interacting species. As a consequence of the nearest-neighbor interaction, even species that possess well-adapted abilities, with high barriers, can be undermined in their existence by weak neighbors.

For the special case $M = 1$, we obtain the original Bak-Sneppen model where each species has a single internal degree of freedom to represent its over-all fitness.

It has been shown [9,10], via the "gap" equation, that the sequence of selective updates at extremal sites drives the system from any initial state to a self-organized critical state where species exhibit punctuated equilibrium behavior with bursts of evolutionary activity correlated over all spatial and temporal extents. In this state almost all species have reached fitnesses above a critical threshold, enjoying long periods of quiescence, interrupted by intermittent activity when changes in neighboring species force a readjustment in their own barriers. These generic features are preserved for arbitrary $M$.

To understand the nature of the SOC state, it is useful to consider the case where, at a certain time $s = 0$, the smallest random number in the system has a value $\lambda$. A $\lambda$ avalanche is defined as the collection of barrier values at subsequent times $s > 0$ which are below $\lambda$; the $\lambda$ avalanche that started at $s = 0$ ends at the first instant $s > 0$ when the number of barriers in the system below $\lambda$ vanishes. All barriers that are below the threshold value $\lambda$ at any one instant in time are called "active" because they make up the $\lambda$ avalanche. When the system reaches the state that almost all barriers are evenly distributed above $\lambda = \lambda_c$, the vanishingly small fraction of active levels below $\lambda_c$ form $\lambda_c$ avalanches that are distributed according to power laws in their spatial and temporal extents, i. e. they possess no cutoff. The lack of a cutoff, which leads to divergent expectation values, indicates that a critical state has been reached.

We introduce a cascade mechanism that describes the dynamics of $\lambda$ avalanches for $M \to \infty$. The case $M \to \infty$ of our model is special because the existing active barriers that any species possesses can only be changed if these barriers themselves become the global minimum. While the nearest-neighbor interaction chooses one out of the $M$ barriers at every site next to the minimal site for an update, there are infinitely many barriers *on each site* and no existing active barrier is ever likely to be chosen in this way. To simplify the algebra, a slight modification is made without restricting the generality of the results: At each time step during the avalanche, the smallest active barrier is set to unity instead of being replaced by a new random number.

Now, consider the probability $P_\lambda(s)$ for a $\lambda$ avalanche, which started at time $s = 0$, to end at time $s$. The properties of such an avalanche can be related to smaller avalanches by considering the state of the system after one update. Clearly, $P_\lambda(0) = 0$. First, we examine the one dimensional case. The avalanche ends at $s = 1$ only if the initial active barrier places two new barriers above $\lambda$. This happens with probability $(1-\lambda)^2$, so that $P_\lambda(1) = (1-\lambda)^2$. For $s \geq 2$,

$$P_\lambda(s) = 2\lambda(1-\lambda)P_\lambda(s-1) + \lambda^2 \sum_{s'=0}^{s-1} P_\lambda(s')P_\lambda(s-1-s'), \qquad (1)$$

where an avalanche of duration $s$ is obtained in various ways from smaller avalanches that are initiated after the first update. If exactly one new active barrier is created, with probability $\lambda(1-\lambda)$, an avalanche of duration $s$ is obtained by following the first update with an avalanche of duration $s - 1$. If two new barriers are created, with probability $\lambda^2$, two avalanches ensue. Both of these avalanches evolve in a *statistically independent* manner for $M \to \infty$. Since only one of these avalanches can be updated at each time step, their combined duration has to add up to $s - 1$ for this process to contribute to the avalanche of duration $s$. Thus, we simply need to sum over all possible products of two avalanches of combined duration $s - 1$.

A generating function $p(x) \equiv \sum_{s=0}^{\infty} P_\lambda(s) x^s$ with

$$p(x) = \left[1 - \sqrt{1-4\lambda(1-\lambda)x}\right]^2 \left(4\lambda^2 x\right)^{-1} \qquad (2)$$

solves Eqs. (1) [15], and we find

$$P_\lambda(s) = \frac{(1-\lambda)\Gamma(s+1/2)}{\lambda \Gamma(1/2)\Gamma(s+2)} \left[4\lambda(1-\lambda)\right]^s. \qquad (3)$$

A critical point exists for $\lambda = \lambda_c = 1/2$. Near $\lambda_c$ the distribution of avalanche sizes has the scaling form

$$P_\lambda(s) \sim s^{-3/2} G\left(s(\Delta\lambda)^2\right), \quad \Delta\lambda = \lambda_c - \lambda. \qquad (4)$$

The average size of an avalanche is given by $\langle s \rangle = \sum_s s P_\lambda(s) = p'(1)$. The divergence close to the critical state defines the critical exponent $\gamma$: $\langle s \rangle \sim (\Delta\lambda)^{-\gamma}$ with $\gamma = 1$. It is easy to see that Eq. (4) is not changed in higher dimensions [15]. Our results for the temporal behavior agree with the exact results in Refs. [12] where the dynamics of a random neighbor (infinite range) model was solved using different methods. That model, though, does not possess any spatial correlations or punctuated equilibrium behavior.

For our model, we can use the same mechanism to solve for the spatial correlations in the critical state. Again, consider a $\lambda$ avalanche initiated at time $s = 0$ at the origin ($r = 0$). For the one dimensional model, we define $N_\lambda(r)$ as the probability that the $\lambda$ avalanche that ensues will never have a minimum at a particular site of distance $r$ away from the origin, before the avalanche terminates. Due to the initial state, $N_\lambda(0) = 0$. If no new active barriers are created in the first update, the $\lambda$ avalanche ends and will not spread to distances $r > 0$. If a single new active barrier is created to either side of the origin, then the probability for $N_\lambda(r)$ is related to $N_\lambda(r+1)$ and $N_\lambda(r-1)$, respectively. If two active barriers are created, two avalanches ensue that evolve independently. Thus, the probability that neither one spreads to site $r$ is the product of their individual probabilities. Then, for $r \leq 1$,

$$N_\lambda(r) = (1-\lambda)^2 + \lambda(1-\lambda)\left[N_\lambda(r-1) + N_\lambda(r+1)\right] + \lambda^2 N_\lambda(r-1) N_\lambda(r+1). \qquad (5)$$

When $r \to \infty$, $N_\lambda(r) \to 1$ since any given avalanche cannot spread to an infinitely distant site. Defining $N_\lambda(r) = 1 - f_r$, we find

$$f_{r+1} - 2f_r + f_{r-1} = \left(\frac{1}{\lambda} - 2\right) f_r + \lambda f_{r-1} f_{r+1} \qquad . \quad (6)$$

For thresholds below the critical value, $f_r$ falls exponentially fast for large $r$. The nonlinear difference equation (6) can be solved exactly at the critical point:

$$f_r = \frac{12}{(r+3)(r+4)} \qquad \text{for} \qquad \lambda = \lambda_c = \frac{1}{2}. \quad (7)$$

Close to the critical point, this quantity also obeys a scaling form: $f_r \sim 1/r^2 H(r(\Delta\lambda)^{1/2})$ for large $r$.

Since only avalanches of spatial extent larger than $r$ can contribute to $f_r$, the probability to have an avalanche of total spatial extent exactly of $r$ is $P_{\lambda_c}(r) \sim r^{-\tau_R}$ with $\tau_R = 3$. Numerical measurements [16] of $f_r$ are in perfect agreement with the exact result in Eq. (7) and a simulation of $P_{\lambda_c}$ gave $\tau_R = 3.0 \pm 0.1$ for one dimension. We also find $\tau_R = 3$ in higher dimensions. There, the equation corresponding to Eq. (6) is asymptotically dominated by $\partial_r^2$ in the Laplacian, and by the quadratic nonlinearity.

In the SOC state, spatial and temporal correlations are profoundly interrelated. This interrelation is expressed via scaling relations. In a broad class of SOC models, the knowledge of just two scaling coefficients, such as $\tau$ and $\tau_R$, is sufficient to determine any other known coefficient of the SOC state, including the approach to the attractor, through these scaling relations [10]. For example, the activity in the SOC state spreads in a subdiffusive manner, $r \sim s^{1/D}$, where $D$ is the avalanche dimension. Normalization of probability requires that $\tau_R - 1 = D(\tau - 1)$, so $D = 4$ for the $M \to \infty$ model. Fig. shows that numerical calculations confirm our analytical result for $D$. In fact, this exponent can be calculated directly, without resorting to scaling relations, from a more general recursion relation for the probability $N_\lambda(r, s)$ that a $\lambda$ avalanche of duration $s$ does not spread to a particular site at distance $r$. In Ref. [17] we will discuss this quantity and show that it contains both Eqs. (1) and (5) as special cases.

In the SOC state, the distribution of distances between subsequent minimal sites scales as a power law $P_{\text{jump}}(r) \sim r^{-\pi}$ for large $r$ [7]. Its exponent is obtained through $\pi = 1 + D(2 - \tau)$ [10], i.e. $\pi = 3$ in the $M \to \infty$ model. We find $\pi \simeq 3.03 \pm 0.08$ in simulations involving $\approx 10^{10}$ updates of the one dimensional model.

In a long-lived avalanche, each site is visited many times, leading to punctuated equilibrium behavior. The intervals between subsequent returns to a given site are analogous to periods of stasis for a given species. As shown in Fig. , the accumulated number of returns to a given site forms a "Devil's staircase"; the plateaus in the staircase are the periods of stasis for that species. The punctuations, i. e. the times when the number of returns increases, occupy a vanishingly small fraction of the time scale on which the evolutionary activity proceeds. The distribution of plateau sizes is the same as the distribution of first returns of the activity to a given site, $P_{\text{FIRST}}(s)$. It has been found that $P_{\text{FIRST}}(s) \sim s^{-\tau_{\text{FIRST}}}$ for large $s$ with $\tau_{\text{FIRST}} = 2 - d/D$ [10]. For $M \to \infty$, $\tau$ and $\tau_R$, and hence $D$, do not change with dimension $d$, and it is $\tau_{\text{FIRST}} = 2 - d/4$ for all $d \leq 4$. Thus, for $d = 1$ we predict $\tau_{\text{FIRST}} = 7/4$. Numerically, we find $\tau_{\text{FIRST}} = 1.73 \pm 0.05$. We present numerical calculations in agreement with the exact result for $\tau_{\text{FIRST}}$, and other features of the $M \to \infty$ model, in $d > 1$ elsewhere [17].

Exact results for individual scaling coefficients help to separate models of SOC, and the phenomena they represent, into different universality classes. For instance, comparison of the exact results for the $M \to \infty$ model with the numerically obtained scaling coefficients for the $M=1$ Bak-Sneppen model shows that they belong to different universality classes. Numerical results suggest that any finite $M$ model crosses over at large scales to $M = 1$ behavior. The robustness of the model for SOC and punctuated equilibrium behavior with respect to changing internal parameters indicates that these features may underlie the dynamical behavior of more complicated systems in nature. For instance, the $M \to \infty$ limit of the model can be used to show that many changes in the microscopic rules do not affect the universality class.

The cascade mechanism we used to derive the distribution functions for spatio-temporal correlations of the evolutionary activity has some similarity to the path integral approach used in the Fixed Scale Transformation method [18]. For $M \to \infty$, we explicitly calculate the statistical weight of each configuration in terms of the sum over all histories that lead to the particular configuration. It is straightforward to generalize the cascade mechanism to the case of finite $M$, although not to solve it. Taking into account interactions between active barriers in Eq. (5) gives $N_\lambda(0) = 0$ and, for $r \geq 1$,

$$N_\lambda(r) = (1-\lambda)^2 + \lambda(1-\lambda) \left[ N_\lambda(r-1) + N_\lambda(r+1) \right]$$
$$+ \lambda^2 \langle \hat{N}_\lambda(r-1) \hat{N}_\lambda(r+1) \rangle. \quad (8)$$

Here, $\langle \hat{N}_\lambda(r-1) \hat{N}_\lambda(r+1) \rangle$ is the joint probability distribution function for an avalanche that has two active barriers, each one step to the left and to the right of the origin, to never spread to $r$ before it terminates. In $M \to \infty$, this two-point correlation function factors because there is no interference between avalanches. For finite $M$, however, this two-point correlation function must be determined by the next equation in the hierarchy that includes all possible evolutions up to two update steps. It is straightforward to deduce this equation and show that it will introduce three-point (and eventually higher) correlation functions. We do not currently know how to solve the resulting cascade hierarchy in a systematic way. Our exact results suggest that introducing many internal degrees of freedom per site may also be useful in studying other models for nonequilibrium phenomena.

This work was supported by the U. S. Department of Energy under Contract No. DE-AC02-76-CH00016. MP thanks the U.S. Department of Energy Distinguished Postdoctoral Fellowship Program for financial support.

---

Plot of the mean-square distance $\sqrt{\langle r^2 \rangle}$ covered by the activity in a $\lambda_c$ avalanche as a function of time $s$. The mean-square distance was calculated using the probability distribution in space and time for the location of the minimal barrier in avalanches that are initiated at the origin at $s = 0$. The probability distribution was sampled by evolving $10^6$ such avalanches up to $s = 800$. To emphasize that the distribution asymptotically scales as $r/s^{1/D}$ with $D = 4$, we have rescaled the mean-square distance by $s^{-1/4}$.

Punctuated equilibrium behavior for the evolution of a single species in the one-dimensional $M \to \infty$ model. The vertical axis is the total number of returns of the activity to site 100 as a function of time $s$. Note the presence of plateaus (periods of stasis) of all sizes. The distribution of plateau sizes scales as $s^{-7/4}$.